\documentclass[3p,times]{elsarticle}
\usepackage{ecrc}
\volume{00}

\firstpage{1}

\journalname{Nuclear Physics A}

\runauth{A.~Mart\'inez Torres et al.}

\jid{NUPHA}

\jnltitlelogo{Nuclear Physics A}

\CopyrightLine{2012}{Published by Elsevier Ltd.}
\usepackage{graphicx}
\usepackage{amsmath}
\usepackage{amsfonts}
\usepackage{amssymb}

\begin{document}

\begin{frontmatter}

\dochead{}

\title{Three-body hadron systems with strangeness}

\author[a]{A. Mart\'inez Torres}\author[a]{K. P. Khemchandani}\author[b]{D. Jido}\author[c]{Y. Kanada-En'yo}\author[d]{E. Oset}

\address[a]{Instituto de F\'isica, Universidade de S\~ao Paulo, C.P 66318, 05314-970 S\~ao Paulo, SP, Brazil.}
\address[b]{Yukawa Institute for Theoretical Physics, Kyoto University,
Kyoto 606-8502, Japan.}
\address[c]{Department of Physics, Kyoto University, Kyoto 606-8502, Japan.}
\address[d]{Departamento de F\'isica Te\'orica and IFIC, Centro Mixto Universidad de 
Valencia-CSIC,\\
Institutos de Investigaci\'on de Paterna, Aptdo. 22085, 46071 Valencia,
Spain.}

\begin{abstract}
Recently, many efforts are being put in studying three-hadron systems made of mesons and baryons and interesting results are being found.
In this talk, I summarize the main features of the formalism used to study such three hadron systems with strangeness $S=-1,0$ within a framework built on the basis of unitary chiral theories and solution of the Faddeev equations. In particular, I present the results obtained for the $\pi\overline{K}N$, $K\overline{K}N$ and $KK\overline{K}$ systems and their respective coupled channels. In the first case, we find four $\Sigma$'s and two $\Lambda$'s with spin-parity $J^P=1/2^+$, in the $1500-1800$ MeV region, as  two meson-one baryon $s$-wave resonances. In the second case, a $1/2^+$ $N^*$ around 1900 MeV is found. For the last one a kaon close to 1420 MeV is formed, which can be identified with $K(1460)$.
\end{abstract}
\end{frontmatter}

\section{Introduction}
In the conventional quark model, hadrons can be classified in two groups on the basis of their quark content: baryons, constituted by three quarks, and mesons, formed by a quark-antiquark pair. Keeping this idea in mind, it was just a matter of time to build up the different baryons  and mesons observed in nature within quark models. All one needed was to write all the possible combinations of three quarks or quark-antiquark pairs as a function of their charge and strangeness. Radial and orbital excitations of these quarks from the ground state to different high energy levels could  generate more hadrons having a relatively short lifetime, which are called resonances.

However, this picture of the hadron resonances seems to be too simple to understand the properties of the different states present in nature. For example, the lowest excited state found of the nucleon is the $N^*(1440)$. However, in a three-quark model for a baryon, the state $N^*(1535)$ is expected to be the first excitation of the nucleon, with a radial excitation of a quark. From the kinematical point of view, this implies providing an energy of around $600$ MeV to one of the three quarks in the nucleon.  This energy is sufficient to create, for example, a pion or two pions or an eta meson.  Therefore, it is plausible to think that to describe the properties of states as the ones mentioned above, the interaction of a pion and an nucleon,
or an eta and a nucleon could be more crucial than the quark structure of the state.

Therefore, a theory which uses the hadrons as the degrees of freedom instead of quarks can be more useful to understand the properties of some of the meson and baryon resonances found in nature.
This situation is in fact very much expected when one studies interactions of hadrons in a low and an intermediate energy region. Here, due to the inherent confinement of the quarks, these can not be considered as the asymptotic states of the theory.

Having this idea in mind, in the last years, non-perturbative unitarity techniques based on chiral Lagrangians  have been applied to the study of different meson-baryon and meson-meson systems, for example, $\overline{K}N$, $\pi\Sigma$, $\pi N$, $K\overline{K}$, etc., and dynamical generation of many baryon and meson resonances has been found. For instance,  $\Lambda(1405)$, $\Lambda(1520)$, $N^*(1535)$, etc., in the baryon sector and $f_{0}(980)$, $a_{0}(980)$, $f_0(500)$, etc., in the meson sector \cite{oller,jido,nieves,borasoy}.

Since the interaction in such two-body systems is strongly attractive, the addition of one more meson or baryon could further lead to the generation of new states in which case the interaction of the three hadrons could be determinant in understanding some of the experimental findings.

With the motivation to search for such states and to understand their properties, we have extended the well studied two-body chiral formalism  to  three-body systems. Here, I summarize the method developed for the investigation of three hadron systems as well as the results found for the generation of resonances and/or bound states in $\pi\overline{K} N$, $K\overline{K} N$, and $K K\overline{K}$ systems.

\section{Formalism}
A rigorous study of three-body systems requires solving the Faddeev equations \cite{faddeev}.~In this formalism, the $T$ matrix of the three-body system is written as sum of three partitions, $T^1$, $T^2$ and $T^3$, i.e.,
 
 \begin{align}
 T=T^1+T^2+T^3.\label{T}
 \end{align}
 Each of these partitions, $T^i$, with $i=1,2,3$, represents that contribution to the total $T$ matrix in which particles  $j$ and $k$, with $i\neq j\neq k$=1,2,3, are the last ones in interacting. Thus, particle $i$ is a spectator in the last interaction. In this way, the summation of the three partitions in Eq.~(\ref{T}) accounts for the different permutations and combinations of the sequence of different pair interactions among the three particles. These partitions satisfy the Faddeev equations:
 \begin{align}
 T^i=t^i+t^iG[T^j+T^k],\label{Fa}
 \end{align}
 where $t^i$ denotes the $t$-matrix for that two body subsystem where particle $i$ is absent and $G$ is the three-body Green's function for the system.  
 
The solution of the Faddeev equations in its exact form is a cumbersome task, due to which one often resorts to approximations.  While most conventional studies of three-body systems  use potentials in the momentum space, usually separable potentials to make the solution of the Faddeev equations feasible,  we used two particle $t$ matrices generated from the solution of the Bethe-Salpeter equation in its on-shell factorization form~\cite{oller,jido}: 
\begin{align}
t^i=(1-V^ig)^{-1}V^i.\label{BS}
\end{align}
In our approach, the kernel $V^i$ present in Eq.~(\ref{BS}) is obtained from the lowest order chiral Lagrangian describing the interaction between the particles  and   
 $g$ is the two-body loop function, divergent in nature and which is regularized using a natural cut-off or subtraction constant when using dimensional regularization~\cite{oller,jido}. Equation~(\ref{BS}) is solved in a coupled channel approach, generating in this way different hadron resonances where the hadron-hadron interaction is essential for understanding the properties of the states found. It is worth mentioning that the availability of more precise data in recent times is helping in constraining the parameters involved in the higher order terms of the chiral Lagrangians, leading to more precise determination of observables~\cite{borasoy, nicola,yoichi}. The contribution of these higher order terms in the three-body calculation should be checked in the future.

Once the $t$ matrices are obtained by solving Eq.~(\ref{BS}), we can proceed to solve Eq.~(\ref{Fa}). The Faddeev equations are integral equations and the $t$ matrices which enter in the equations are off-shell.  The interesting point of using chiral amplitudes for solving Eq.~(\ref{Fa}) is that they can be split into two terms: one which is called on-shell, since it is calculated as a function of the Mandelstam variable $s$ with the external particles on their mass shell, and other off-shell which goes as $q^2-m^2$, with $q$ the momentum and $m$ the mass of the particle, and which vanishes when the external particles are on-shell. We found that when the chiral amplitude is inserted in Eq.~(\ref{Fa}) a cancellation occurs between the contribution arising from the off-shell part of these amplitudes and three hadron contact terms originating from the chiral Lagrangians. This cancellation is exact and analytical in the SU(3) limit as shown in the case of $s$-wave interactions in Refs.~\cite{mko1,mko2,mko3}.  We have further checked that when the condition of the SU(3) limit is removed, the sum of the contributions coming from the off-shell part of the two-body $t$ matrices and three hadron contact terms, which constitutes the sources of three-body forces, turns out to be very small as compared to the one obtained from the on-shell part. Thus one can consider only the on-shell two-body t-matrices and ignore the three-body forces coming from the chiral Lagrangians.  This gives rise to the use of on-shell amplitudes in the solution of the Faddeev equations, which in terms of these on-shell two-body $t$ matrices read as

\begin{equation}
T^{ij}_R=t^ig^{ij}t^j+t^i[G^{iji}T_R^{ji}+G^{ijk}T_R^{jk}]\label{mko}.
\end{equation}
In Eq.~(\ref{mko}) $g^{ij}\equiv G$ is the three-body Green function of the system and $G^{ijk}$ a loop function involving three-hadron propagators (for more details, see Refs.~\cite{mko1,mko2,mko3}).
The Faddeev partitions of Eq.~(\ref{Fa}) are related to the $T^{ij}_R$ partitions through:
\begin{align}
T^i&=t^i\delta^3(\vec{k}_i-\vec{k}^\prime_i)+T_R,\nonumber\\
T_R&\equiv\sum_{i=1}^3\sum_{j\neq i=1}^3 T^{ij}_R,
\end{align}
where $\vec{k}_i$ ($\vec{k}^\prime_i$) is the momentum of the particle $i$ in the initial (final) state.

The two-body $t$ matrices and $G$ functions entering in Eq.~(\ref{mko}) depend on the invariant masses of the corresponding subsystems, $s_{ij}$, and the total energy of the system, $s$. In this way, the $T^{ij}_R$ amplitudes are calculated as a function of $s$ and the invariant mass of one of the pairs, for example the one related to particles 2 and 3, $s_{23}$, since the other kinematic variables can be obtained as a function of these two variables~\cite{mko1,mko2,mko3}.

After solving Eq.~(\ref{mko}) for a certain system, since we work in the charge basis, we project the resulting amplitudes in an isospin base in which the states are labelled by the total isospin of the three-body system, $I$, and the isospin related to one of the subsystems, $I_{sub}$, and search for peaks in the squared amplitude which can be identified with resonances.
\section{Results}
\subsection{The $\pi\overline{K} N$ system and coupled channels.}
One of the successes of unitary chiral dynamics is the reproduction of the $\Lambda(1405)\,S_{01}$ ($J^P=1/2^-$) properties, which has been found to get dynamically generated (with a two pole structure \cite{jido}) from the $\overline{K}N$ interaction and its coupled channels. If another pseudoscalar meson is added to this system, in $S$-wave, it results into states with spin-parity $J^P=1/2^+$. The lightest pseudoscalar meson which can be added is the pion. 
The resulting three-body system would posses a mass $\sim$ 1570 MeV. This is exactly the region where the $1/2^+$ hyperon resonances are poorly understood~\cite{PDG}. The poor status of these low-lying $S=-1$ states is evident from the following facts: a) The spin-parity assignment for many of these states is unknown, e.g., for $\Sigma(1480)$, $\Sigma(1560)$, etc., b) the partial-wave analysis and production experiments have been often archived separately in the PDG listings, e.g., for $\Sigma(1620)$, $\Sigma(1670)$, c) other times, e.g., in case of $\Lambda(1600)$, it is stated that existence of two resonances, in this energy region, is quite possible \cite{PDG}. This situation lead us to question if some of these poorly understood states could be understood as three-hadron resonances. In such a case, it would be difficult to investigate them considering only two hadron decay channels, which would result in poor information related to these resonances. In fact, some of them, like $\Lambda (1600)$, $\Sigma(1660)$,  decay to three-body final states with large branching ratios~\cite{prakhov,prakhov2}, implying that  some of the wave functions of the $1/2^+$ resonances in the $S=-1$ sector have an appreciable two-mesons and one baryon contribution, like $\pi\overline{K} N$, $\pi\pi\Sigma$, $\pi\pi\Lambda$, and $\pi K\Xi$.

Motivated by this, to search for possible three-body states, we started by taking all the combinations of a pseudoscalar meson of the $0^-$ SU(3) octet and a baryon of the $1/2^+$ octet which couple to $S = -1$ with any charge. To this system we add a pion and obtain twenty-two coupled channels with net charge zero: $\pi^0 K^- p$, $\pi^0\overline{K}^0 n$, $\pi^0\pi^0\Sigma^0$, $\pi^0\pi^+\Sigma^-$, $\pi^0\pi^-\Sigma^+$, $\pi^0\pi^0\Lambda$, $\pi^0\eta\Sigma^0$, $\pi^0\eta\Lambda$, $\pi^0 K^+\Xi^-$, $\pi^0 K^0\Xi^0$, $\pi^+ K^- n$, $\pi^+\pi^0\Sigma^-$, $\pi^+\pi^-\Sigma^0$, $\pi^+\pi^-\Lambda$, $\pi^+\eta\Sigma^-$, $\pi^+ K^0\Xi^-$, $\pi^-\overline{K}^0 p$, $\pi^-\pi^0\Sigma^+$, $\pi^-\pi^+\Sigma^0$, $\pi^-\pi^+\Lambda$, $\pi^-\eta\Sigma^+$, $\pi^- K^+ \Xi^0$. 

\begin{figure}[h!]
\centering
\includegraphics[width=\textwidth]{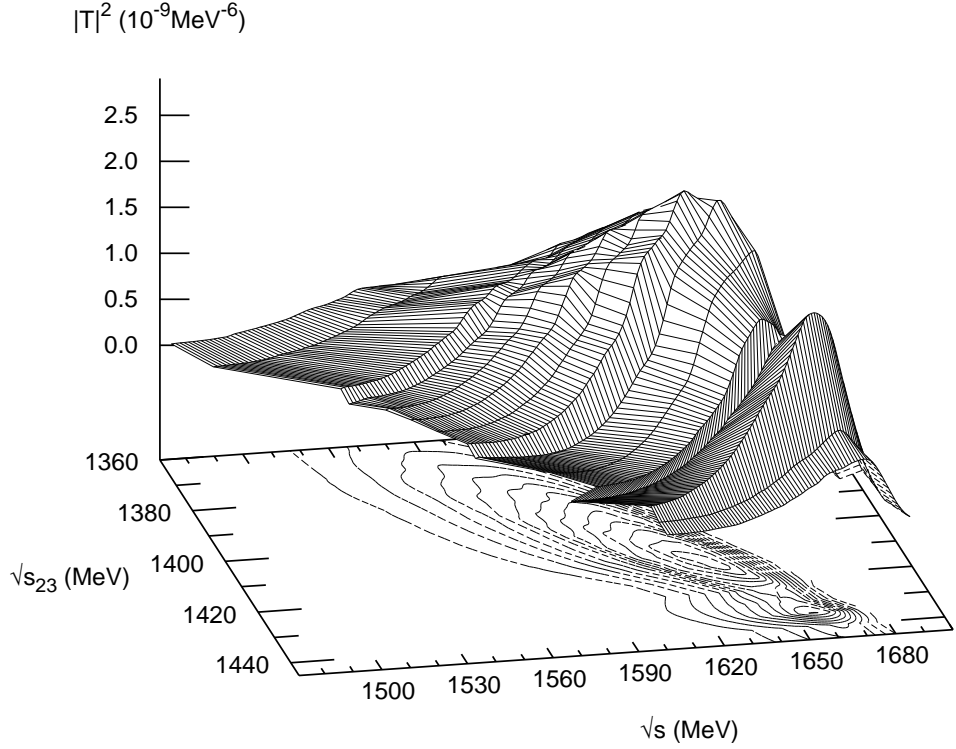}
\caption{Two $\Sigma$ resonances in the  $\pi \pi \Sigma$ amplitude in $I$ = 1, $I_\pi$ = 2 configuration. In the figure  $\sqrt{s}$ is the total energy of the system,  and $\sqrt{s_{23}}$ is the invariant mass of the $\pi\Sigma$ subsystem .}\label{2Sig}
\end{figure}

In Fig.\ref{2Sig}, we show a plot of the squared $T_R$-matrix, calculated within the formalism explained in the previous section, corresponding to the $\pi \pi \Sigma$ channel for total isospin $I$ = 1 with two pions in isospin $I_{\pi\pi}$ = 2. We see two peaks; one at $\sqrt{s}$ = 1656 MeV with $\sim$ 30 MeV of width and another at $\sqrt{s}$ = 1630 MeV with 39 MeV of width. We identify the former peak with the well established $\Sigma(1660 - i100/2)$  \cite{PDG} and predict a $\Sigma(1630)$. A $\Sigma^*$ state with mass around 1630 MeV and $J^\pi=1/2^+$ is not listed by the Particle Data Group (PDG), however, the findings of Refs.~\cite{Armenteros,zou} hint towards its existence. It is interesting to notice that the $\Sigma^*$'s found appear for different values of the $\pi\Sigma$ invariant mass, $\sqrt{s_{23}}$, around 1430 and 1410 MeV, respectively. The observation of these $\Sigma^*$ resonances in a particular experiment could be difficult, since the mass difference between these states is smaller than their respective widths. However, the fact that they are generated for different values of $\sqrt{s_{23}}$ should be helpful in identifying them in a Dalitz plot involving the variables $\sqrt{s}$ and $\sqrt{s_{23}}$.  

In addition, we found some more $\Sigma$ and $\Lambda$ resonances. We summarize our findings in  Table \ref{tablesigma}. As can be seen, we are able to generate all the low-lying $1/2^+$ $\Lambda$ and $\Sigma$ resonances listed by the PDG in the energy region 1500-1800 MeV as two mesons one baryon states.

\begin{table}[h!]
\begin{center}
\begin{tabular}{|c|c|c|c|}
\hline
&$\Gamma$ (PDG)&Peak position (this work)&$\Gamma$ (this work)\\
&(MeV)&(MeV)&(MeV)\\
\hline
\multicolumn{4}{|c|}{Isospin = 1} \\
\hline
$\Sigma(1560)$&10 - 100&1590&70\\
\hline
$\Sigma(1620)$&10 - 100&1630&39\\
\hline
$\Sigma(1660)$&40 - 200&1656&30\\
\hline
$\Sigma(1770)$&60 - 100&1790&2\\
\hline
\multicolumn{4}{|c|}{Isospin = 0} \\
\hline
$\Lambda(1600)$&50 - 250&1568, 1700&60, 136\\
\hline
$\Lambda(1810)$&50 - 250&1740&20\\
\hline
\end{tabular}
\end{center}
\caption{A comparison of the resonances found in our work with the states listed by the PDG.}
 \label{tablesigma}
\end{table}

Here we have limited the discussions to $s$-wave interactions, however, it should be mentioned that a study of the $\pi\overline{K} N$ system including $p$-wave interactions has been done in Ref.~\cite{gal}, where a $3/2^-$ $\Sigma$ resonance of molecular structure is predicted with a mass of 1570 MeV.

\subsection{The $K\overline{K} N$ system}

The $K\overline{K}N$ system was studied in Ref.~\cite{jido2} using effective potentials to describe the interaction
between the different subsystems. As a result, a  $N^{*}$ with $I=1/2$ and $J^{P}$=$1/2^{+}$ was found 
around 1910 MeV  when the $\overline{K}N$ pair forms the $\Lambda(1405)$ 
and, simultaneously, the $K\overline{K}$ pair is resonating as $a_{0}(980)$.  The hadron-hadron distances found in Ref.~\cite{jido2} for the $K\overline{K}N$ state are of about 2 fm, which is as large as typical nucleon-nucleon distance in nuclei.
\begin{figure}[h!]
\centering
\begin{tabular}{cc}
\hspace{-1cm}
\vspace{-0.5cm}
\resizebox{0.5\textwidth}{!}{\includegraphics[width=\textwidth]{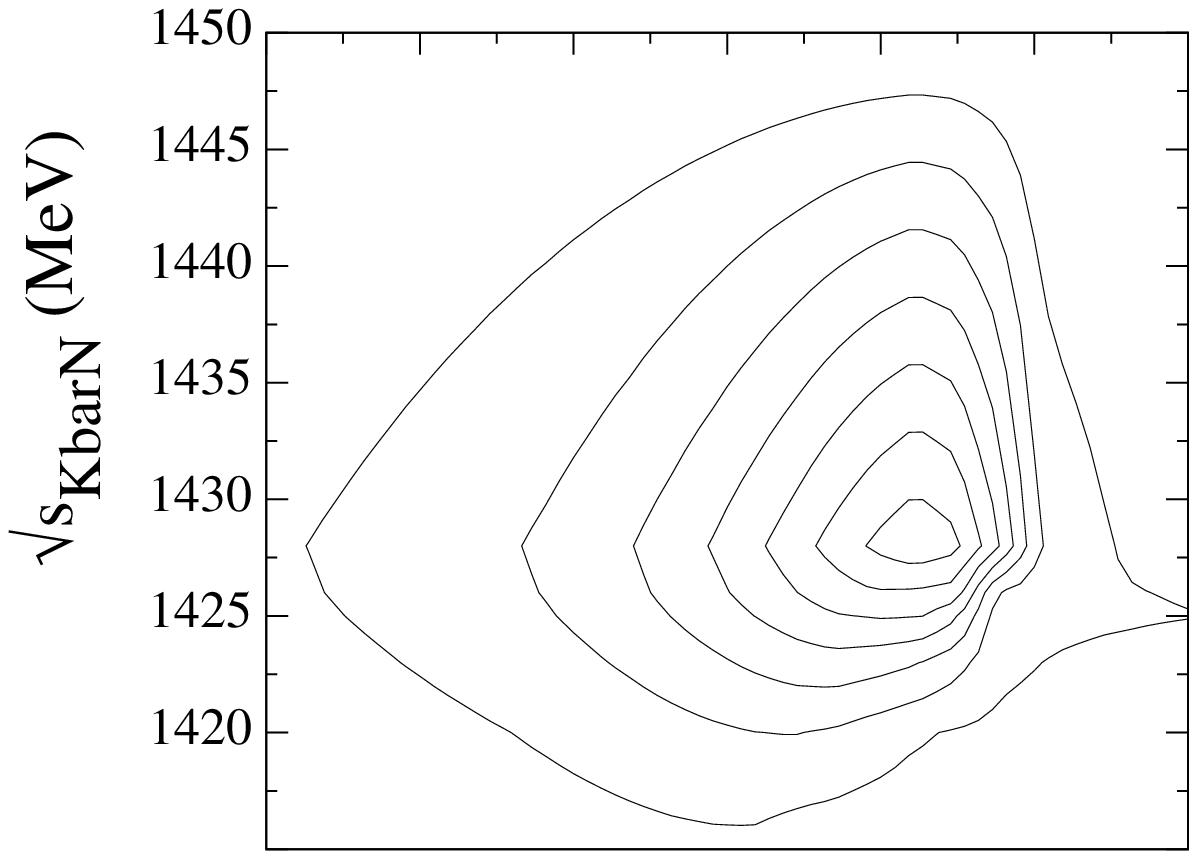}}\\
\hspace{-1cm}
\resizebox{0.5\textwidth}{!}{\includegraphics[width=\textwidth]{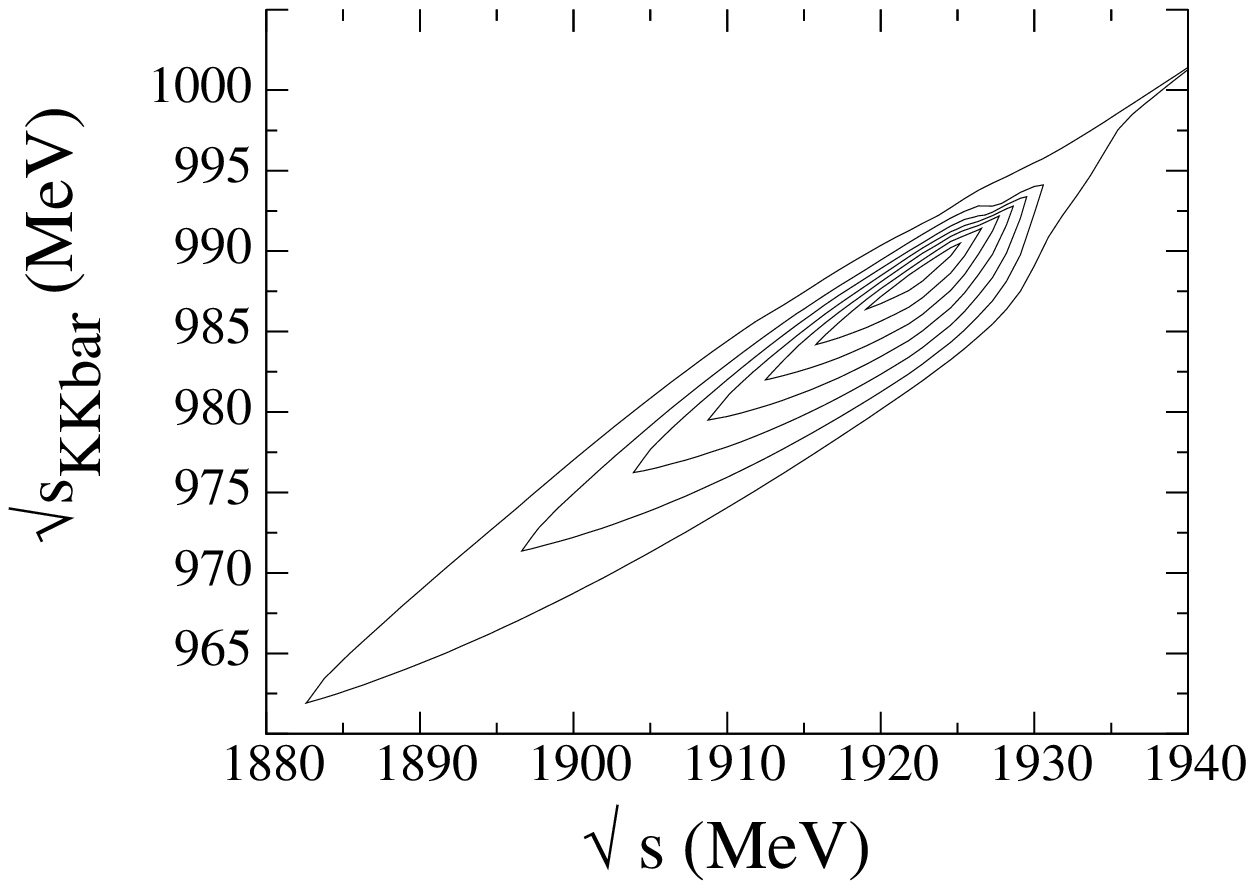}}
\end{tabular}

\caption{Contour plots of the three-body squared amplitude $|T_{R}|^2$ 
for the $N^{*} $ resonance in the $K\overline{K}N$ system
as functions of the total three-body energy, $\sqrt s$, and
the invariant mass of the $\overline{K}N$ subsystem with $I_{\overline{K}N}=0$
(upper panel) or the invariant mass of the $\overline{K}K$ subsystem with 
$I_{\overline{K}K}=1$ (lower panel).} \label{N1}
\end{figure}

To confirm or revoke the findings in Ref.~\cite{jido2}, we solved the Faddeev equations for the $K\overline{K} N$ system and coupled channels~\cite{mko5,mj1}. In Fig.~\ref{N1} we show the contour plots corresponding to the three-dimensional plots of the squared three-body $T_{R}$ matrix for $I=1/2$, $I_{\overline{K}N}=0$ (upper panel) and $I=1/2$, $I_{\overline{K}K}=1$ (lower panel)
plotted as functions of the total energy of the three-body system, $\sqrt{s}$, and the $\overline{K} N$ invariant mass, $\sqrt{s_{\overline{K}N}}$, and the $\overline{K} K$ invariant mass, $\sqrt{s_{\overline{K}K}}$, respectively.  As can be seen in the upper panel, a peak in the squared amplitude is obtained around $\sqrt{s}\sim1922$ MeV when the $\overline{K}N$ subsystem in isospin zero has an invariant mass close to 1428 MeV. In the lower panel, the peak shows up when the invariant mass of the $K\overline{K}$ subsystem is around 987 MeV. Thus, it can be concluded that a $N^*$ resonance with $J^P=1/2^+$ is formed in the $K\overline{K} N$ system when the $\Lambda(1405)$ is generated in the $\overline{K} N$ subsystem and the $a_0(980)$ state is formed in the $K\overline{K}$ subsystem, in agreement with the findings of Ref.~\cite{jido2}.

 A $N^*$ state with these characteristics is not catalogued in the PDG
\cite{PDG}. However, there is a peak in the $\gamma p \to K^+
\Lambda$ reaction at around 1920 MeV, clearly visible in the integrated 
cross section and also at
all angles from forward to backward \cite{saphir,jefflab,mizuki} which could correspond to this state, as suggested in Ref.~\cite{mko4}.
\subsection{The $KK\overline{K}$ system}
Another interesting system to study is the one formed by two kaons and an anti-kaon. In the $K\overline{K}$ system the $f_0(980)$ and $a_0(980)$ are dynamically generated and if the attraction between the
two $K\overline{K}$ pairs is strong enough to overcome the  repulsion between the kaons, a bound state could be formed. 

\begin{figure}[h!]
\centering
\includegraphics[width=0.6\textwidth]{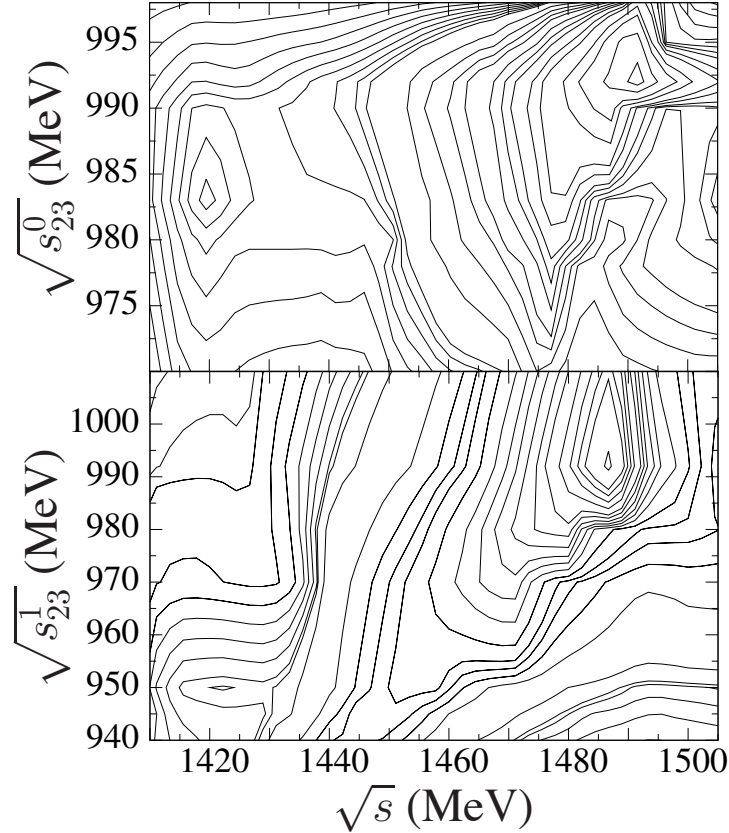}
\caption{Contour plots of the three-body squared amplitude $\Big|T_{R}\Big|^2$
 for the $KK\overline{K} \to KK \overline{K}$ transition 
with total $I=1/2$ as functions of the total three-body energy, $\sqrt s$, and 
the invariant mass of the $K\overline{K}$ subsystem with $I_{23}=0$ (upper panel)
or the invariant mass of the $K\overline{K}$ subsystem with $I_{23}=1$ (lower panel).
} \label{N2}
\end{figure}

With this idea in mind, we perform a Faddeev calculation of the $KK\overline{K}$, $K\pi\pi$ and $K\pi\eta$ systems treating them as coupled channels. In Fig.~\ref{N2}, we show the contour plots associated to  $|T_{R}|^2$ for the process
$KK\overline{K} \to KK\overline{K}$ in the cases  $I=1/2$, $I_{23}=I_{K\overline{K}}=0$ (upper panel) and $I=1/2$, $I_{23}=1$ (lower panel).  First of all,  we see in both panels of Fig.~\ref{N2} a peak structure at an energy  around $3 m_K\sim1488$ MeV (with $m_K$ = 496 MeV the kaon mass) which appears when the invariant masses of the respective $K\overline{K}$ subsystems have a value around $2 m_K$, i.e., their threshold values. If we only considerer $K\pi\pi$ and $K\pi\eta$ as coupled channels, the signal at 1488 MeV is not present. Thus, we conclude that the peak which shows up at 1488 MeV corresponds then to the opening of the three-body $KK\overline{K}$ threshold.  Apart from this, we find a peak at $\sqrt s \sim 1420$ MeV and a width of $\sim$ 50 MeV when $\sqrt {s_{23}} \sim 983$ MeV for the case $I_{23}=0$, as shown in the upper panel of Fig.~\ref{N2}.
As can be seen in the lower panel of Fig.~\ref{N2}, this resonance state also shows up for a value of  $\sqrt {s_{23}}\sim 950$ MeV and $\sqrt s \sim 1420$ MeV when $I_{23}=1$.
Thus, a state at 1420 MeV shows up when the invariant mass of the $K\overline{K}$ pair with isospin zero is close to a value of 983 MeV, implying then that the $f_0(980)$ resonance is formed in the subsystem. However, when the $K\overline{K}$ is projected on $I_{23}=1$, the invariant mass for the $K\overline{K}$ pair has a value around 950 MeV. This value is not exactly in the region where the $a_0(980)$ gets dynamically generated,  but it is also not very far away, and the attraction present in the system helps in forming a three-body bound state. In fact, a recent study of the two body system $Kf_0(980)$ shows the generation of a kaon around 1460 MeV~\cite{alba}.

Following Ref.~\cite{jido2},  we have also studied the $KK\overline{K}$ system using effective potentials to describe the interaction between the pairs of the system. The $\overline{K}K$ interaction strengths were determined  to have a quasibound state with mass 980 MeV and width 60 MeV in isospin 0 and isospin 1, which correspond to
the $f_{0}(980)$ and $a_{0}(980)$ resonances, respectively. This means that the attractive $\overline{K}K$ interactions have the same strengths for both $I_{\overline{K}K}=0$ and $I_{\overline{K}K}=1$.
The strength of the repulsive $KK$ interaction in $I_{KK}=1$ was fixed to reproduce the scattering length $a_{K^{+}K^{+}}=-0.14$, which has been obtained 
from a lattice QCD calculation~\cite{Beane:2007uh}. In the potential model, the three-body wave function is also obtained. 
With the wave function we can investigate the spacial structure of the three-body 
quasibound state. We obtain the root mean squared radius of the $KK\overline{K}$ quasibound state 
to be 1.6 fm. This value is  similar  to the one found for the $K\overline{K}N$ system~\cite{jido2}, which was 1.7 fm.
The average $K$-$K$ distance and the distance 
between the $KK$ cluster and $\overline{K}$ are calculated and found to be 2.8 fm and 
1.7 fm, respectively. 
The distance of the repulsive $KK$ is also very similar to the corresponding 
result for the $KN$ distance in the $K\overline{K}N$ system. 

The kaonic state obtained within the two methods can probably be associated to the $K(1460)$ listed by the PDG \cite{PDG} (which is omitted from the summary table) and  observed  in $K\pi\pi$ partial wave analysis. However, the poor experimental information available in this energy region for kaonic states suggests  that more experiments are needed to confirm the existence of this state.
We get as a result a quasibound state of the $KK\overline{K}$ system
with 21 MeV binding energy and 110 MeV width.
This state appears for an energy similar 
to the one of the resonance obtained in the Faddeev calculation. It is interesting to notice that although the two methods 
are very different, the energy position of the quasibound $KK\overline{K}$ state does not differ very much.
Note, however, that in the potential model used we consider only the single $KK\overline{K}$ channel and
do not take into account the possible modification of the two-body interaction in the 
presence of the third particle. In such simple calculation, for weakly bound systems, 
the resulting binding energy and width correspond to the sum of the binding energies and widths 
of the two-body subsystems, $f_{0}(980)$ and $a_{0}(980)$ in the present case, as discussed in Ref.~\cite{jido2}.

\section{Summary}
In this talk I have presented an approach for solving the Faddeev equations based on unitarized chiral theories, due to which a cancellation between the off shell part of the two body scattering matrices and three body contact terms stemming from the same theory is found.  As a consequence on-shell two body $t$ matrices can be used as input to the Faddeev equations. Within this
formalism, I have shown the results found in three systems and their respective coupled channels, obtaining the generation of several resonances. It is thus interesting to continue with such studies of three hadron systems.
\section*{Acknowledgments}

This work is partly supported by the Spanish Ministerio de Econom\'ia y Competitividad and european  FEDER funds under
the contract FIS2011-28853-C02-01, and the Generalitat Valenciana in the program Prometeo, 2009/090. We acknowledge the support of the European Community-Research Infrastructure Integrating Activity Study of Strongly Interacting Matter (acronym HadronPhysics3, Grant Agreement n.~283286) under the Seventh Framework Programme of the EU. This work is supported in part by the Grant for Scientific Research (No.~22105507 and No.~22540275) from 
MEXT of Japan. A part of this work was done in the Yukawa International Project for  Quark-Hadron Sciences (YIPQS). A. M. T and K. P. K would like to thank the Brazilian funding agencies FAPESP and CNPq for the financial support.

\end{document}